\newcounter{myctr}
\def\myitem{\refstepcounter{myctr}\bibfont\noindent\ifnum\themyctr>9\else\phantom{0}\fi\hangindent17pt\themyctr.\enskip}

\documentclass{ws-ijqi}

\begin{document}

\markboth{Authors' Names}
{Instructions for Typing Manuscripts (Paper's Title)}

\catchline{}{}{}{}{}

\title{Some applications of uncertainty relations in quantum information}

\author{A. S. Majumdar}

\address{S. N. Bose National Centre for Basic Sciences\\
Block JD, Sector III, Salt Lake, Kolkata 700098, India\\
archan@bose.res.in}

\author{T. Pramanik}
\address{LTCI, T\'{e}l\'{e}com ParisTech\\
 23 avenue dItalie, 75214 Paris CEDEX 13, France\\
Tanumoy.Pramanik@telecom-paristech.fr}

\maketitle

\begin{history}
\received{Day Month Year}
\revised{Day Month Year}
\end{history}

\begin{abstract}
We discuss some applications of various versions of uncertainty relations for 
both discrete and continuous variables in the context of quantum information
theory. The Heisenberg uncertainty relation
enables demonstration of the EPR paradox. Entropic uncertainty relations
are used to reveal quantum steering for non-Gaussian continuous variable
states. Entropic uncertainty relations for discrete variables are studied
in the context of quantum memory where fine-graining yields the optimum
lower bound of uncertainty. The fine-grained uncertainty relation is used
to obtain connections between uncertainty and the nonlocality of retrieval
games for bipartite and tipartite systems. The Robertson-Schrodinger uncertainty
relation is applied for distinguishing pure and mixed states of discrete
variables. 
\end{abstract}


\section{Introduction}	


The uncertainty principle is a central feature of quantum mechanics, 
prohibiting certain properties of 
quantum systems from being simultaneously well-defined.
The Heisenberg uncertainty 
relation$^1$ 
lower bounds the product of uncertainties, i.e., the spread 
measured by standard deviation, of measurement outcomes for two 
non-commutating observables$^2$. 
 An improved form of
the uncertainty relation was proposed by Robertson$^3$ 
and Schrodinger$^4$, 
incorporating both commutators and 
anti-commutators of more general observables.  Motivated by
various physical considerations, several other versions
of the uncertainty principle have since been suggested. 
Notable
among them are  reformulations that
take into account the inevitable noise and disturbance associated with 
measurements$^5$. 

Efforts for eliminating the
state-dependence of the lower bound of uncertainty have lead to  the 
formulation of various entropic versions of
the uncertainty principle$^{6,7,8,9}$. 
Entropic
uncertainty relations have been tightened due different effects, such as
the presence of correlations$^{10,11,12,13,14}$. 
A fine-grained version of the uncertainty relation arises as a result of
distinguishing the uncertainty
of obtaining specific combinations of outcomes
 for different measurements$^{15}$. 
 An optimal
lower bound of entropic uncertainty in the presence of any type of
correlations may
be determined by fine-graining$^{16}$. 
For a recent review of uncertainty relations, see Ref.~$^{17}$. 

The subject of quantum information science that has seen  rapid
progress in recent years, was inspired originally to a great extent by
the pioneering work of Einstein, Podolsky and Rosen (EPR)$^{18}$. 
The word `entanglement’ was first coined by Schrodinger
to describe the property of spatially separated but correlated particles
whose paradoxical features were highlighted by EPR. The first
testable formulation of the EPR paradox was proposed$^{19}$ 
using the 
position-momentum uncertainty relation,
in terms of an inequality involving products of inferred variances of 
incompatible
observables. This lead to the experimental realization$^{20}$ 
of the EPR 
paradox  
for the case of two spatially separated and correlated light modes.
A modern formulation of the EPR-Schrodinger concept of quantum steering
based on violations of steering inequalities$^{21}$, 
akin to the Bell-type local-realist inequalities$^{22,23}$, 
is derived
again using uncertainty relations in their entropic version. Entropic
steering relations are indispensable for demonstrating steering in
certain continuous variable systems where correlations are not manifest
up to second order (variances of observables), as shown recently
for several non-Gaussian states$^{24}$. 

Several other important applications of uncertainty relations
in the realm of quantum information processing have been uncovered
in recent years.  The uncertainty principle has been used for discrimination 
between separable and entangled quantum states in the realm of
continuous variable systems$^{25}$.  
The utility of the Robertson-Schrodinger uncertainty relation$^{3,4}$ 
has also been exploited in this 
context$^{26,27}$. 
Moreover, the Robertson-Schrodinger uncertainty relation$^{3,4}$ has
recently been employed in the domain of discrete variables to distinguish 
between pure and mixed states of single as well as bipartite qubit and qutrit
systems$^{28}$. 
The fine-grained uncertainty relation can be used to determine the nonlocality
of the underlying physical system$^{15}$, as has been demonstrated for
the case of bipartite$^{15}$ and tripartite$^{29}$ systems, 
as well as in the arena of biased nonlocal games$^{30}$. 
The uncertainty principle plays a crucial role in the domain of
quantum cryptography since
security of quantum key distribution protocols relies basically on
quantum uncertainty$^{31}$. 
Specifically, the amount of key extractable per
state has been linked to the lower limit of entropic 
uncertainty$^{10,32}$. 

Uncertainty relation in their different versions have many important
applications in quantum information theory. In the present article, we 
review some aspects of a few of these applications, limited mainly
by the areas in the which the present authors have worked upon. 
The plan of this article is as follows. In
the next Section we discuss the Robertson-Schrodinger uncertainty relation
and briefly sketch how it could be used for distinguishing pure states
from mixed states of discrete variables. In Section III we focus on
the topic of quantum steering where steering using the Heisenberg
uncertainty relation as well as entropic steering relations are discussed
in the context of continuous variables. The connection between uncertainty
and nonlocality of quantum games is presented in Section IV
as an application of the fine-grained steering relation. Section V contains
a brief review of entropic uncertainty relations in the presence of
quantum memory. Certain concluding remarks are made in Section VI.

\section{Determining purity of states using the Robertson-Schrodinger uncertainty relation}

In experimental protocols for information processing, the interaction with the 
environment inevitably affects the purity of a quantum system. A relevant
issue for an experimenter is to ascertain whether a prepared pure state
has remained isolated from environmental interaction.
It becomes important to test whether a given quantum state is pure, in 
order to use it effectively as a resource for quantum information 
processing.
The purity of a given state is also related to the 
entanglement of a larger multipartite system of which it may be a 
part$^{33}$. 
The mixedness of states can be quantified by their linear entropy,
which is  a nonlinear functional of the quantum state. The linear entropy
can be extracted from the given state by tomography  which 
usually is expensive in terms of resources and measurements involved.

In this section we discuss how the Robertson-Schrodinger (RS) 
uncertainty relation
may be used to determine the 
mixedness of quantum states of discrete variables. 
For the case of continuous variable 
systems  there exist certain pure states for which the uncertainty as 
quantified by the Robertson-Schrodinger uncertainty relation  is 
minimized$^{34}$, 
The connection of purity
with observable quantities of the relevant states 
have been found$^{35}$. 
It has been
shown recently that the RS uncertainty relation can be used 
to distinguish between pure and
mixed states of finite dimensional systems$^{28}$.

The RS uncertainty relation could be used as 
a witness of mixedness in the following way.
For any pair of observables $A,B$ and for any quantum state represented by 
the density operator $\rho$, the RS uncertainty relation 
can be written as$^{3,4}$ 
\begin{eqnarray}
Q(A,B,\rho) \ge 0
\label{gur1}
\end{eqnarray}
where
\begin{eqnarray}
Q(A,B,\rho)&=&(\Delta A)^2 (\Delta B)^2- |\frac{\langle[A,B]\rangle}{2}|^2 \nonumber \\
&& - |(\frac{\langle\{A,B\}\rangle}{2}-\langle A\rangle \langle B\rangle)|^2
\label{gur2}
\end{eqnarray}
with $(\Delta A)^2$ and $(\Delta B)^2$ representing the variances of the 
observables, $A$ and $B$, respectively, given by
$(\Delta A)^2=(\langle A^2\rangle)-(\langle A\rangle)^2$, $(\Delta B)^2=(\langle B^2\rangle)-(\langle B\rangle)^2$, and the square (curly) brackets representing
the standard commutators (anti-commutators) of the corresponding operators.
The quantity $Q(A,B,\rho)$ involves the measurable quantities, i.e., the 
expectation values and variances of the relevant observables in the state 
$\rho$. States of a $d$-level quantum system are in one to one correspondence 
with Hermitian, positive semi-definite, unit trace operators acting on a 
$d$-dimensional Hilbert space. The defining properties of these density 
operators $\rho$ are (i) $\rho\dagger=\rho$, (ii) $\rho\geq 0  $, (iii) 
$tr[\rho]=1$. Pure states correspond to the further condition $\rho^2=\rho$ 
which is equivalent to the scalar condition $tr[\rho^2]=1 $. Hence, complement 
of the trace condition can be taken as a measure of mixedness given by
the linear entropy defined for a $d$-level system as 
\begin{eqnarray}
S_l(\rho) = (\frac{d}{d-1})(1-tr(\rho^{2}))
\label{linentrop}
\end{eqnarray}
We now describe how the  quantity $Q(A,B,\rho)$ can act as an 
experimentally realizable measure of mixedness of a system$^{28}$. 

 Let us  here
discuss the case of
two-level systems. 
 The density operator for qubit systems can be expressed 
in terms of the 
Pauli matrices.
The state of a single qubit can be written as
$\rho(\vec{n})=\frac{(I+\vec{n}.\vec{\sigma})}{2},   \>\>    \vec{n}\in \mathbb{R}^{3}$.
Positivity of this Hermitian unit trace matrix demands $ |\vec{n}|^2\leqslant1$.
It follows that single qubit states are in one to one correspondence with the 
points on or inside the closed unit ball  centred at the origin of 
$\mathbb{R}^{3}$. Points on the boundary correspond to pure states.  For 
a pair of 
suitably chosen spin observables,  the RS relation is satisfied as an 
equality for the 
states extremal, i.e., the pure states, and as an inequality for points other 
than extremals, i.e., for the mixed states$^{28}$. 
The linear entropy of the state $\rho$ can be written as
$S_l(\rho)=(1-\vec{n}^{2})$.
If we choose spin observables along two different directions, i.e., 
$A=\hat{r}.\vec{\sigma}$ and $B=\hat{t}.\vec{\sigma}$, then  $Q$ becomes
\begin{eqnarray}
Q(A,B,\rho)=(1-(\Sigma r_{i}t_{i})^{2})S_l(\rho)
\label{qub}
\end{eqnarray}
It thus follows that for $\hat{r}.\hat{t}=0$,  $Q$ coincides 
with  the linear entropy. For orthogonal spin measurements, the 
uncertainty quantified by the RS relation, $Q$ and the linear entropy $S_l$ are exactly 
same for single qubit systems. Thus, it turns out that $Q=0$ is both a 
necessary and sufficient condition for any single qubit system to be pure when 
the pair of observables are qubit spins along two different directions.

For two-qubit systems the states considered
may be taken to be
polarized along a specific known direction, say, the $z$-
axis forming the Schmidt decomposition basis. In order to enable 
$Q(A,B,\rho)$ to be a 
 mixedness measure,  $A$ and $B$ are chosen 
for the two-qubit case to be of the form
$A=(\hat{m}.\vec{\sigma}^{1})\otimes(\hat{n}.\vec{\sigma}^{2})$, and
$B=(\hat{p}.\vec{\sigma}^{1})\otimes(\hat{q}.\vec{\sigma}^{2})$, respectively,
where $\hat{m},\hat{n},\hat{p},\hat{q}$ are unit vectors.
For enabling $Z(A,B,\rho)$ to be used
for determining the purity of the given two qubit
state, the appropriate choice of observables 
$A$ and $B$ is found to be that of lying on the two
dimensional $x-y$ plane (i.e.,$\hat{m},\hat{n},\hat{p},\hat{q}$ are all taken
to be on the $x-y$ plane), normal to the $z$-axis pertaining
to the relevant Schmidt decomposition basis.  Then, $Q(A,B,\rho)=0$ 
 necessarily holds good for
pure two-qubit states whose individual spin orientations
are all along a given direction (say, the $z$-axis) normal to
which lies the plane on which the observables $A$ and $B$
are defined. On the other hand, $Q(A,B,\rho) > 0$ holds good for most settings 
of $A$ and
$B$ for two qubit isotropic states, for the Werner class of states given by
$\rho_{w}=((1-p)/4)I+p\rho_{s}$
 ($\rho_s$ is the two-qubit singlet state),
as well for other types of one parameter two-qubit states which comprise of 
pure states whose
individual spin orientations are all along the same given
direction normal to the plane on which the observables
$A$ and $B$ are defined.
 
The RS uncertainty relation has been shown to determine the purity of
qutrit systems, as well$^{28}$. Three-level systems are of
fundamental relevance in laser physics, and have generated much recent 
interest from the perspective
of information processing$^{36}$. 
It has been shown using 
examples of single
and bipartite class of qutrit states that the RS uncertainty relation 
can be satisfied as an equality
for pure states while it remains an inequality for mixed states by
 the choice of suitable observables.  An observational scheme 
 which can detect mixedness of qutrit systems unambiguously,
requires less resources compared to tomography, and is 
implementable through the measurement of Hermitian
 witness-like operators$^{28}$. It may be relevant to note here though
that the set of pure states is not convex, and hence, such witness-like 
operators do not arise from any geometrical separability criterion
inherent to the theory of entanglement witnesses$^{37}$, 
that has been
applied more recently to the cases of teleportation witnesses$^{38}$, 
as well as for witnesses of absolutely separable states$^{39}$. 

The operational determination of purity using the RS relation requires
a few additional steps.
A scheme for using the uncertainty 
relation to determine whether a given state is pure or mixed, provided
the prior knowledge of the basis is available, has been outlined in Ref.$^{28}$.
The limitation of instrumental precision 
could  make the 
observed value of $Q$ for
pure states to be a small number in stead of exactly zero. In order to
take into account the experimental inaccuracy, a parameter  
$\varepsilon$ is introduced in the analysis. For a single-qubit system, 
by choosing the measurement
settings for  $A$ and $B$ as qubit spins along $z$ and $x$ 
directions, respectively,   the measured value of the uncertainty  
obtained as $Q\ge\varepsilon$ leads to the conclusion that  
the given state is mixed. This prescription
of determining mixedness holds for all single-qubit states 
$\rho(\vec{n})=\frac{(I+\vec{n}.\vec{\sigma})}{2}$, except those lying
in the narrow range $1 \ge n \ge \sqrt{1- 2 \varepsilon/3}$, as determined by 
putting $Q < \varepsilon$.

To summarize, the RS uncertainty
relation  is able to distinguish between
pure and mixed states for a broad category of two- and three-level systems.
 For single party systems, the scheme works for all qubits
and up to three-parameter family of qutrit states$^{40}$. 
For bipartite systems, the scheme has been shown to work
for the mixture of two arbitrary pure states, the isotropic class, and the
Werner class of states, as well.
The determination of mixedness using GUR may
require in certain cases a considerably lesser number of measurements compared 
to tomography.
In the case of single  
qutrit states, full tomography involves the estimation of eight parameters,
while through the prescription detailed in Ref.$^{28}$ sometimes four  
measurements may suffice for
 detecting purity of a single qutrit state.
A maximum of eight measurements 
suffices to distinguish between pure and mixed states of single qutrit up 
to three-parameter families.  The difference in the number of required measurements is substantially
enhanced for composite states. For two qubits, the RS relation requires up to 
five measurements
compared to fifteen required by tomography. 
For the case of two-qutrits 
the measurement of at most eight expectation
values  suffices.

\section{Quantum steering}

The Einstein-Podolsky-Rosen (EPR) paradox$^{18}$ has not only inspired
a huge body subsequent debate, but has played a pivotal role in the unfolding 
of several rich features of quantum mechanics relevant for information
processing.  Considering a position-momentum correlated state
of two particles, and assuming the notions of spatial separability, locality,
and reality to hold true at the level of quantum particles, EPR argued that
that the quantum mechanical description of the state of a particle is not
complete. The
EPR paradox arises from the correlations between two non-commuting
observables of a sub-system with those of the other sub-system, for instance,
the correlations between the measurement outcomes
of positions and momenta for two separated particles, i.e.,
$<x,p_y> \neq 0$, with $<x>=0=<p_y>$ individually. 
 Due to the presence of correlations,
the measurement of the position of, say, the first particle leads one to infer 
the correlated value of
the position for the second particle (say, $x_{\inf}$). Now, if the momentum 
of the second particle is measured
giving the outcome, say $p$, the value of the product of uncertainties $(\Delta x_{\inf})^2 (\Delta p_{\inf})^2$ may
 turn out to be lesser than that allowed by the uncertainty principle, {\it viz.} $(\Delta x)^2 (\Delta p )^2 \ge 1$, thus leading to the paradox.

Following the work of EPR, Schrodinger$^{41}$ 
observed that  correlations between spatially separated
particles entailed the possibility of
steering of the state on one side merely  by the choice
of the measurement basis on the other side, without in any way having direct
access to the affected particle. The word 'entanglement' was first coined
by Schrodinger to describe the property of such spatially separated but
correlated particles.  Consider a bipartite entangled
state which may be expressed in two different ways, as
\begin{eqnarray}
\vert\Psi\rangle = \sum_{n=1}^{\infty}c_n\vert\psi_n\rangle\vert u_n\rangle =
 \sum_{n=1}^{\infty}d_n\vert\phi_n\rangle\vert v_n\rangle
\label{ensemb}
\end{eqnarray}
where $\{\vert u_n\rangle\}$ and $\{\vert v_n\rangle\}$ are two orthonormal
bases for one of the parties (say, Alice). If Alice chose to measure in the 
$\{\vert u_n\rangle\}$  ($\{\vert v_n\rangle\}$) basis,  she 
 projects
Bob's system into one of the states $\vert\psi_n\rangle$ ($\vert\phi_n\rangle$).
Note that though there is no physical interaction between Alice and Bob,
 the ensemble of $\vert\psi_n\rangle$s is in general different from the ensemble
of $\vert\phi_n\rangle$s.
This ability of Alice to affect Bob's state due to her choice of the 
measurement basis was dubbed as ``steering'' by Schrodinger$^{41}$. 

A testable formulation of the EPR 
paradox was proposed many years later by Reid$^{19}$ 
for continuous variable systems using the position-momentum uncertainty relation. An 
inequality involving products of inferred variances of incompatible 
observables was derived in the context of continuous variables, as follows. 
Consider  the quadrature phase components of two correlated and spatially
separated light fields.  The quadrature amplitudes associated with the fields $E_{\gamma}=C[\hat{\gamma} e^{-i\omega_{\gamma} t} + \hat{\gamma}^{\dagger} e^{i\omega_{\gamma} t}]$ (where, $\gamma\in\{a,b\}$, are the bosonic operators for two different modes, $\omega_{\gamma}$ is the frequency, and
$C$ is a constant incorporating spatial factors taken to be equal for each mode) are given by
\begin{eqnarray}
\hat{X}_{\theta}=\frac{\hat{a}e^{- i \theta} + \hat{a}^{\dagger} e^{i \theta}}{\sqrt{2}},
\hspace{0.5cm}
\hat{Y}_{\phi}=\frac{\hat{b}e^{- i \phi} + \hat{b}^{\dagger} e^{i \phi}}{\sqrt{2}},
\label{Quard}
\end{eqnarray}
where,
\begin{eqnarray}
\hat{a} &=& \frac{\hat{X} + i \hat{P}_x}{\sqrt{2}},\hspace{0.5cm} \hat{a}^\dagger = \frac{\hat{X} -i \hat{P}_x}{\sqrt{2}},\nonumber\\
\hat{b}&=&  \frac{\hat{Y}+i \hat{P}_y}{\sqrt{2}}, \hspace{0.5cm} \hat{b}^\dagger = \frac{\hat{Y}- i \hat{P}_y}{\sqrt{2}},
\label{boson_op}
\end{eqnarray}
and the commutation relations of the bosonic operators are given by $[\hat{a},\hat{a}^{\dagger}]=1=[\hat{b},\hat{b}^{\dagger}]$.
The correlations between the quadrature amplitudes $\hat{X}_{\theta}$ and $\hat{Y}_{\phi}$ are defined by the correlation coefficient, $ C_{\theta,\phi}$  as$^{19,20}$ 
\begin{eqnarray}
C_{\theta,\phi}=\frac{\langle \hat{X}_{\theta} \hat{Y}_{\phi} \rangle}{\sqrt{\langle \hat{X}^2_{\theta} \rangle \langle \hat{Y}^2_{\phi}  \rangle}},
\label{Cr_f}
\end{eqnarray}
where $\langle \hat{X}_{\theta} \rangle=0=\langle \hat{Y}_{\phi} \rangle$. The correlation is perfect for some values of $\theta$ and $\phi$, if $|C_{\theta,\phi}|=1$, and vanishes  for uncorrelated variables.

As a consequence of correlations in the measurement outcomes, the quadrature amplitude $\hat{X}_{\theta}$ can be inferred by measuring the corresponding amplitude $\hat{Y}_{\phi}$.  In realistic situations the correlations are not perfect because of the interaction with the environment as well as finite detector efficiency. Hence, the estimated amplitudes $\hat{X}_{\theta 1}$ and $\hat{X}_{\theta 2}$ with the help of $\hat{Y}_{\phi 1}$ and $\hat{Y}_{\phi 2}$, respectively, are subject to inference errors, and given by$^{19}$ 
\begin{eqnarray}
\hat{X}_{\theta 1}^{e}=g_1 \hat{Y}_{\phi 1},
\hspace{0.5cm}
\hat{X}_{\theta 2}^{e}=g_2 \hat{Y}_{\phi 2},
\label{Est}
\end{eqnarray}
where $g_1$ and $g_2$ are scaling parameters.
Now, one may choose $g_1$, $g_2$, $\phi 1$, and $\phi 2$ in such a way that $\hat{X}_{\theta 1}$ and $\hat{X}_{\theta 2}$ are inferred with the highest possible accuracy. The errors given by the deviation of the estimated amplitudes from the true amplitudes $\hat{X}_{\theta 1}$ and $\hat{X}_{\theta 2}$ are captured by $(\hat{X}_{\theta 1}- \hat{X}_{\theta 1}^{e})$ and $(\hat{X}_{\theta 2}- \hat{X}_{\theta 2}^{e})$, respectively. The average errors of the inferences are given by
\begin{eqnarray}
(\Delta_{\inf} \hat{X}_{\theta 1})^2 &=& \langle (\hat{X}_{\theta 1}- \hat{X}_{\theta 1}^{e})^2\rangle =  \langle (\hat{X}_{\theta 1}- g_1 \hat{Y}_{\phi 1})^2\rangle, \nonumber \\
(\Delta_{\inf} \hat{X}_{\theta 2})^2 &=& \langle (\hat{X}_{\theta 2}- \hat{X}_{\theta 2}^{e})^2\rangle =  \langle (\hat{X}_{\theta 2}- g_2 \hat{Y}_{\phi 2})^2\rangle.
\label{Erro}
\end{eqnarray}
The values of the scaling parameters $g_1$ and $g_2$ are chosen such that
$\frac{\partial (\Delta_{\inf} \hat{X}_{\theta 1})^2}{\partial g_1} =0 = \frac{\partial (\Delta_{\inf} \hat{X}_{\theta 2})^2}{\partial g_2}$, from which it follows that
\begin{eqnarray}
g_1 = \frac{\langle \hat{X}_{\theta 1} \hat{Y}_{\phi 1} \rangle}{\langle \hat{Y}_{\phi 1}^2 \rangle},
\hspace{0.5cm}
g_2 = \frac{\langle \hat{X}_{\theta 2} \hat{Y}_{\phi 2} \rangle}{\langle \hat{Y}_{\phi 2}^2 \rangle}.
\label{g's}
\end{eqnarray}
The values of $\phi 1$ ($\phi2$) are obtained by maximizing $C_{\theta1,\phi1}$ ($C_{\theta 2,\phi2}$).
 Now, due to the commutation relations $[\hat{X},\hat{P}_X]=i;~~[\hat{Y},\hat{P}_Y]=i$, it is required
 that the product
 of the variances of the above inferences $(\Delta_{\inf} \hat{X}_{\theta 1})^2 (\Delta_{\inf} \hat{X}_{\theta 2})^2 \ge 1/4$. Hence, the EPR paradox occurs if the correlations in the field quadratures lead to
 the condition
 \begin{eqnarray}
EPR \equiv (\Delta_{\inf} \hat{X}_{\theta 1})^2 (\Delta_{\inf} \hat{X}_{\theta 2})^2 < \frac{1}{4}.
\label{P_Uncer}
\end{eqnarray}

Experimental realization of the EPR paradox was first carried out 
by Ou et al.$^{20}$ 
using two spatially separated and correlated light modes. Similar
demonstrations of the EPR paradox using quadrature amplitudes of other 
radiation fields were performed later$^{42}$. 
Subsequent works have shown that
the Reid inequality is effective in demonstrating the EPR paradox for
systems in which correlations appear at the level of variances, though
there exist
several pure entangled states which do not display steering through the
Reid criterion.
Moreover, in systems
with correlations manifesting in higher than the second moment, the Reid
formulation generally fails to show occurrence  of the EPR paradox, even though
Bell nonlocality may be exhibited$^{43,44}$. 

On the other hand, a modern formulation of quantum steering in terms
of an information theoretic task was proposed by the 
work of Wiseman et al.$^{21,45}$. 
They considered a bipartite situation in which
 one of two parties (Alice) prepares a  quantum state
and sends one of the particles to Bob. The procedure is repeated as many
times as required. Bob's particle is assumed to possess a definite
state, even if it is unknown to him (local hidden state). No such 
assumption is made for Alice, and hence, this
formulation of steering is an asymmetric task. 
Alice and Bob make measurements on their respective 
particles, and communicate classically. Alice's task is to convince Bob
that the state they share is entangled. If correlations between Bob's 
measurement results and Alice's  declared results can be explained by 
a local hidden state (LHS) model for Bob, he is not convinced. This is
because Alice could have drawn a pure state at random from some ensemble and 
sent it to Bob, and then chosen her result based on her knowledge of this LHS.
 Conversely, if the correlations cannot be so explained, then the state must 
be entangled. Alice will be successful in her task of steering if she can 
create genuinely different ensembles for Bob by steering Bob's state. 

Using similar formulations for entanglement
as well as Bell nonlocality, a clear distinction between these three types
of correlations is possible using joint probability distributions, 
with entanglement being the
weakest, steering the intermediate, and Bell violation the strongest
of the three. Bell nonlocal states constitute a strict subset of steerable
states which, in turn, are a strict subset of entangled states. For the case 
of pure entangled
states of two qubits the three classes overlap. 
An experimental
demonstration of these differences has been performed for mixed
entangled states of two qubits$^{46}$.

 For the case of continuous variables, Walborn et al.$^{43}$ 
have
proposed another steering
condition which is derived using the the entropic uncertainty 
relation (EUR)$^{6}$.
EUR for the position and momentum distribution of a quantum system is
 given by
\begin{eqnarray}
h_Q(X)+h_Q(P)\geq \ln \pi e.
\label{entropy_uncertainty}
\end{eqnarray}
 Walborn et al.$^{43}$  considered a joint probability
distribution of two parties corresponding to a non-steerable state for
which there exists a local hidden state (LHS) description, given by
\begin{eqnarray}
\mathcal{P}(r_A,r_B)=\sum_\lambda \mathcal{P}(\lambda)\mathcal{P}(r_A|\lambda)\mathcal{P}_Q(r_B|\lambda),
\label{steer2}
\end{eqnarray}
where, $ r_A $ and $ r_B $ are the outcomes of measurements $ R_A $ and $ R_B $ respectively;   $ \lambda $ are hidden variables that specify an ensemble of 
states; $ \mathcal{P} $ are general probability distributions; and $ \mathcal{P}_Q $ are probability distributions corresponding to the quantum state specified by $ \lambda $. Now, using a rule for conditional probabilities
$P(a,b|c) = P(b|c)P(a|b)$ which holds when $\{b\} \in \{c\}$, i.e., there 
exists a local hidden state of Bob predetermined by Alice, it follows that
the conditional probability $\mathcal{P}(r_B| r_A)$ is given by
\begin{eqnarray}
\mathcal{P}(r_B|r_A)=\sum_\lambda \mathcal{P}(r_B,\lambda|r_A)
\label{steer3}
\end{eqnarray}
with $P(r_B,\lambda | r_A) = P(\lambda |r_A)P_Q(r_B|\lambda)$. Note that
(\ref{steer2}) and (\ref{steer3}) are similar conditions for 
non-steerability. Next, considering the relative entropy (defined for two
distributions $p(X)$ and $q(X)$ as $\mathcal{H}(p(X)||q(X))= \sum_xp_x\ln(p_x/q_x)$) between the
probability distributions $ \mathcal{P}(r_B,\lambda|r_A) $ and $ \mathcal{P}(\lambda|r_A)\mathcal{P}(r_B|r_A) $ , it follows from the positivity of
relative entropy that
\begin{eqnarray}
\sum_\lambda \int dr_B \mathcal{P}(r_B,\lambda|r_A) \ln \frac{\mathcal{P}(r_B,\lambda|r_A)}{\mathcal{P}(\lambda|r_A)\mathcal{P}(r_B|r_A)}\geq 0
\end{eqnarray}
Using the non-steering condition (\ref{steer3}), the definition of the 
conditional entropy ($h(X|Y) = -\sum_{x,y} p(x,y)\ln p(x|y)$), and averaging over all measurement 
outcomes $r_A$, it
follows that the conditional entropy $h(R_B|R_A)$ satisfies
\begin{eqnarray}
h(R_B|R_A) \ge \sum_{\lambda} \mathcal{P}(\lambda) h_Q(R_B|\lambda)
\label{cond1}
\end{eqnarray}
Considering a pair of variables $S_A,S_B$ conjugate to $R_A,R_B$, a similar
bound on the conditional entropy may be written as
\begin{eqnarray}
h(S_B|S_A) \ge \sum_{\lambda} \mathcal{P}(\lambda) h_Q(S_B|\lambda)
\label{cond2_Steer}
\end{eqnarray}
For the LHS model for Bob, note that the entropic uncertainty relation 
(\ref{entropy_uncertainty}) holds for each state marked by $\lambda$. 
Averaging over all hidden variables, it follows that
\begin{eqnarray}
 \sum_{\lambda} \mathcal{P}(\lambda)\biggl(h_Q(R_B|\lambda)
+ h_Q(S_B|\lambda)\biggr) \ge \ln \pi e
\label{cond3}
\end{eqnarray} 
Now, using the bounds (\ref{cond1}) and (\ref{cond2_Steer}) in the relation
(\ref{cond3}) one gets the entropic steering inequality given by
\begin{eqnarray}
h(R_B|R_A)+h(S_B|S_A)\geq \ln \pi e.
\label{entropy_steering}
\end{eqnarray}
Entropic functions by definition incorporate correlations 
up to all orders, and  the Reid criterion  follows as a
limiting case of the entropic steering relation$^{43}$. 

EPR steering for Gaussian as well as non-Gaussian states has been studied 
in the literature$^{43,24,47}$.  Non-Gaussian states may be generated 
by the process of photon subtraction and addition$^{48}$, 
and these 
states generally have higher degree of entanglement than the Gaussian states. 
We conclude this section by discussing the example of steering by one
such non-Gaussian state, {\it viz.}, the eigenstate of the two-dimensional
harmonic oscillator.   The energy eigenfunctions of
the two-dimensional harmonic oscillator may be expressed in terms of
Hermite-Gaussian (HG) functions given by$^{48}$ 
\begin{eqnarray}
u_{nm}(x,y) &&= \sqrt{\frac{2}{\pi}} \left(\frac{1}{2^{n+m} w^2 n!m!}\right)^{1/2} \nonumber\\
&& \times H_n \left(\frac{\sqrt{2}x}{w}\right) H_m \left(\frac{\sqrt{2}y}{w}\right) e^{-\frac{(x^2+y^2)}{w^2}},
\nonumber \\
 \int |u_{nm}(x,y)|^2 dx dy &&=1
\label{hermite}
\end{eqnarray}
Entangled states may be constructed  from superpositions of
 HG wave functions
\begin{eqnarray}
\Phi_{nm}(\rho,\theta) = \sum_{k=0}^{n+m} u_{n+m-k,k}(x,y)\frac{f_k^{(n,m)}}{k!}(\sqrt{-1})^k \nonumber \\
\times \sqrt{\frac{k! (n+m-k)!}{n! m! 2^{n+m}}}
\label{legherm}
\end{eqnarray}
\begin{eqnarray}
f_k^{(n,m)} = \frac{d^k}{dt^k} ((1-t)^n(1+t)^m)|_{t=0},
\label{def11}
\end{eqnarray}
where $\Phi_{nm}(\rho,\theta)$, the Laguerre-Gaussian (LG) functions
 are  given by$^{48}$
\begin{eqnarray}
\Phi_{nm}(\rho,\theta) = e^{i(n-m)\theta}e^{-\rho^2/w^2}(-1)^{\mathrm{min}(n,m)}
\left(\frac{\rho \sqrt{2}}{w}\right)^{|n-m|} 
\label{waveLG} \\
\times \sqrt{\frac{2}{\pi n! m ! w^2}}
L^{|n-m|}_{\mathrm{min}(n,m)} \left(\frac{2\rho^2}{w^2}\right) (\mathrm{min}(n,m)) ! \nonumber
\end{eqnarray}
with
 $\int |\Phi_{nm}(\rho,\theta)|^2 dx dy =1$,
where $w$ is the beam waist, and $L_p^l(x)$ is the generalized Laguerre polynomial. 
The superposition (\ref{legherm}) is like a Schmidt decomposition
thereby
signifying the entanglement of the LG wave functions.

In terms of dimensionless quadratures
$\{X,~P_X\}$ and $\{Y,~P_Y\}$, given by
$x (y) \rightarrow \frac{w}{\sqrt{2}} ~~X (Y)$, and
$p_x (p_y) \rightarrow \frac{\sqrt{2} \hbar}{w} ~~P_X (P_Y)$,
the canonical commutation relations are $[\hat{X},\hat{P}_X]=i;~~[\hat{Y},\hat{P}_Y]=i$, and the operator $\hat{P}_X$ and $\hat{P}_Y$ are given by
$\hat{P}_X = - i \frac{\partial}{\partial X}$ and $\hat{P}_Y = - i \frac{\partial}{\partial Y}$, respectively. The Wigner function corresponding to the LG wave function
  in terms of the scaled variables is given by$^{24}$
\begin{eqnarray}
W_{nm}(X,P_X;Y,P_Y)&=&\frac{(-1)^{n+m}}{(\pi)^{2}} L_{n}[4(Q_0+Q_2)]
\label{WF_LG_n1_n2} \\
&& L_{m}[4(Q_0-Q_2)]~exp(-4Q_0) \nonumber
\end{eqnarray}
where
$Q_0  =  \frac{1}{4}\left[ X^2 + Y^2 + P_X^2+P_Y^2\right]$, and 
$Q_2  =   \frac{XP_Y-YP_X}{2}$.
It was shown in Ref.$^{24}$ that the Reid criterion is unable to reveal steering
for the LG wave function. The entropic steering inequality in this case may
be written in terms of the conjugate pairs of dimensionless quadratures, (\ref{entropy_steering}) given by
\begin{eqnarray}
h(\mathcal{X}|\mathcal{P_Y})+h(\mathcal{P_X|}\mathcal{Y})\geq \ln \pi e,
\label{LG_entropy_steering}
\end{eqnarray}
where $ X,~Y,~P_X $ and $ P_Y $ are the outcomes of measurements $ \mathcal{X},~\mathcal{Y},~\mathcal{P_X} $ and $ \mathcal{P_Y} $ respectively.
For $ n=0 $ and $ m=0 $, the LG wave function factorizes into a product
state with the corresponding Wigner function given by
\begin{eqnarray}
W_{00}(X,P_X;Y,P_Y)=\frac{e^{-X^2-Y^2-P_X^2-P_Y^2}}{\pi^2}.
\end{eqnarray}
In this case the relevant entropies turn out to be
$h(\mathcal{X},\mathcal{P_Y})=h(\mathcal{P_X},\mathcal{Y})=\ln \pi e$ and 
$h(\mathcal{Y})=h(\mathcal{P_Y})=\frac{1}{2}\ln \pi e$, and hence, the
entropic steering inequality becomes saturated$^{24}$, i.e.,
\begin{eqnarray}
h(\mathcal{X}|\mathcal{P_Y})+h(\mathcal{P_X|}\mathcal{Y}) = \ln \pi e.
\end{eqnarray}
For $ n=1 $ and $ m=0 $, the Wigner function has the form
\begin{eqnarray}
W_{10}(X,P_X;Y,P_Y)
&=& e^{-X^2-Y^2-P_X^2-P_Y^2}  \\
&& \times \frac{(P_X - Y)^2 +(P_Y+X)^2 -1}{\pi^2} \nonumber
\end{eqnarray}
and the relevant entropies are given by
$h(\mathcal{X},\mathcal{P_Y})=h(\mathcal{P_X},\mathcal{Y}) \approx 2.41509$,
and $h(\mathcal{Y})=h(\mathcal{P_Y}) \approx1.38774$. Hence, the entropic
steering relation in this case becomes
\begin{eqnarray}
h(\mathcal{X}|\mathcal{P_Y})+h(\mathcal{P_X|}\mathcal{Y}) \approx 2.05471 < \ln \pi e 
\end{eqnarray}
Steering is thus demonstrated here.  For higher values of angular
momentum, the violation of the inequality
becomes stronger for higher values of $n$, as shown in Ref.$^{24}$.

It may be noted that the Laguerre-Gaussian functions
are physically realizable field configurations$^{49}$ 
with interesting topological$^{50}$ 
and coherence$^{51}$ 
properties, and are considered to be potentially useful for several
information processing applications$^{52}$. 
Steering has been demonstrated using the entropic steering relation for
other classes of non-Gaussian states such as photon subtracted squeezed
vaccum states and N$00$N states in Ref.$^{24}$ where it has been proposed that
it may be easier to detect entaglement in some such states using steering
compared to the manifestation of Bell violation.
Note also that further
generalizations
of entropic steering inequalities to the case of symmetric 
steering$^{53}$, 
loss-tolerant steering$^{54}$, 
as
well as to the case of steering with qauntum memories$^{55}$ 
have
also been proposed recently.

\section{Fine-graining and its connection with nonlocality}

Uncertainty relations impose restrictions on the knowledge about the properties
of a system described by its state of a system. The Heisenberg uncertainty 
relation prohibits the certain prediction of the measurement outcomes of two 
non-commutating observables. For example, when one predicts certainly the spin 
orientation of a qubit along the $z$-axis, the knowledge of spin orientation 
of that qubit along the $x$-axis is completely uncertain, as the probability 
of getting spin up and down are equal. 
With the motivation of distinguishing the uncertainty inherent in
obtaining any combination of outcomes for
different measurements, Oppenheim and Wehner$^{15}$  proposed a fine-grained 
form of the uncertainty relation. Such fine-graining is aimed at capturing 
the plurality of
simultaneous possible outcomes of a set of measurements. Considering bipartite
systems they formulated a fine-grained uncertainty relation for a special class
of nonlocal retreival games for which there exist only one winning answer for
one of the two parties. The upper bound of the uncertainty relation which is
also the maximum winning probability of the retrieval game was shown to
specify the degree of nonlocality of the underlying physical
theory. In particular, such an upper bound is applicable to discriminate
between the degree of nonlocality pertaining to classical theory, quantum
theory, and no-signalling theory with maximum 
nonlocality for bipartite systems$^{15}$. Similar formulations of fine-graining
in the context of nonlocal games have been later used to distinguish the 
nonlocality of tripartite systems$^{29}$, as well as in the context of biased 
bipartite and tripartite games$^{30}$.

The fine-grained uncertainty relation (or rather, a set of relations) as proposed
by Oppenheim and Wehner$^{15}$ is given by
\begin{eqnarray}
P(\sigma ,\textbf{x}):= \displaystyle\sum_{t=1}^n p(t) p(x^{(t)}|t)_{\sigma} \leq \zeta_{\textbf{x}}(\mathcal{T},\mathcal{D})
\label{FUR1}
\end{eqnarray}
where $P(\sigma ,\textbf{x})$ is the  probability of possible outcomes written as a string $\textbf{x}=\{x^{(1)}, ..., x^{(n)}\}$ corresponding to a set of measurements $\{t\}$ $(\in \mathcal{T})$ chosen with probabilities $\{p(t)\}$ ($\in \mathcal{D}$, the probability distribution of choosing measurements), $p(x^{(t)}|t)_{\sigma}$ is the probability of obtaining outcome $x^{(t)}$ by performing measurement labeled `t' on the state of a general physical system $\sigma$, $n (=|\mathcal{T}|)$ is the total number of different measurement  settings, and $\zeta_{\textbf{x}}(\mathcal{T},\mathcal{D})$ is given by
\begin{eqnarray}
\zeta_{\textbf{x}}(\mathcal{T},\mathcal{D})= \max_{\sigma} \displaystyle\sum_{t=1}^n p(t) p(x^{(t)}|t)_{\sigma}
\label{maxfur}
\end{eqnarray}
where the maximum is taken over all possible states allowed on a particular system. The uncertainty of measurement outcome occurs for the value of $\zeta_{\textbf{x}}(\mathcal{T},\mathcal{D})<1$. The value of $\zeta_{\textbf{x}}(\mathcal{T},\mathcal{D})$ is bound by the particular physical theory. The no-signaling theory with maximum nonlocality gives the upper bound $\zeta_{\textbf{x}}(\mathcal{T},\mathcal{D})=1$.   For  the case of
the single qubit in quantum theory, the form of the fine-grained uncertainty 
relation is given by
\begin{eqnarray}
P(\mathcal{T},\sigma_A)=\displaystyle\sum_{t=1}^n p(t) p(a=x^{(t)}|t)_{\sigma_A}\leq \zeta_{\textbf{x}}(\mathcal{T},\mathcal{D})
\end{eqnarray}
where $p(a=x^{(t)}|t)_{\sigma_A}=Tr[A_t^a.\sigma_A]$
with $A_t^a$  being the measurement operator corresponding to measurement setting `t' giving outcome `a', and $\zeta_{\textbf{x}}(\mathcal{T},\mathcal{D})=\max_{\sigma_A} P(\mathcal{T},\sigma_A)$.
Here the maximum is taken over all possible single qubit states.
The value of $\zeta_{\textbf{x}}(\mathcal{T},\mathcal{D})$ that occurs for the spin measurements along the z-axis  and along the x-axis  with equal probability ( $p(t)=1/2$) on the eigenstates of $(\sigma_x+\sigma_z)/\sqrt{2}$ and $(\sigma_x-\sigma_z)/\sqrt{2}$, is $(\frac{1}{2} + \frac{1}{2 \sqrt{2}})$.
 
The connection between  fine-graining  and
nonlocality was observed by Oppenheim and Wehner$^{15}$ for the
case of bipartite systems. They provided specific examples of 
nonlocal retrieval games (for which there exist only one winning answer for
one of the two parties) for the purpose of discriminating different types
of theories by the upper bound of $\zeta$ (the degree of nonlocality). 
According to these games, Alice and Bob receive questions `s' and `t' respectively, with some probability distribution $p(s,t)$ (for simplicity, $p(s,t)=p(s) p(t)$); and their answer `a' or `b' will be winning answers determined by the 
set of rules, i.e., for every setting `s' and the corresponding outcome `a' of 
Alice, there is a string $\textbf{x}_{s,a}=(x_{s,a}^{(1)}, ..., x_{s,a}^{(n)})$ of 
length $n=|\mathcal{T}|$ that determines the correct answer $b=x_{s,a}^{t}$ for 
the question `t' for Bob. In the particular game considered, 
Alice and Bob share a state $\rho_{AB}$ which is emitted and distributed by a 
source. Alice and Bob are spatially separated enough so that no signal can 
travel while experimenting. Alice performs either of her measurements $A_{0}$ 
and $A_{1}$ and Bob, either of $B_{0}$ and $B_{1}$ at a time. These measurements 
having the outcomes $+1$ and $-1$, can be chosen by Alice and Bob without 
depending on the choice made by the other. The CHSH inequality$^{23}$ 
\begin{equation}
\frac{1}{4} [E(A_{0}B_{0})+ E(A_{0}B_{1})+E(A_{1}B_{0})-E(A_{1}B_{1})]\leq \frac{1}{2}
\end{equation}
holds for any local hidden variable model and can be violated when measurements 
are done on quantum particles prepared in entangled states. Here $E(A_{i}B_{j})$
 are the averages of the product of measurement outcomes of Alice and Bob with $i,j=0,1$.

In the context of the above game, Alice and Bob 
receive respective binary questions $s,t \in \{0,1\}$ (i.e., representing two different measurement settings on each side), and they win the game if their 
respective outcomes (binary) $a,b\in \{0,1\}$ satisfy the condition $a\oplus b=s.t$. At the starting of the game, Alice and Bob discuss their strategy (i.e., choice of shared bipartite state and also measurement). They are not allowed to
communicate with each other once the game has started. The probability of 
winning the game for a physical theory described by bipartite state ($\sigma_{AB}$) is given by
\begin{eqnarray}
P^{game}(\mathcal{S},\mathcal{T},\sigma_{AB})=\displaystyle\sum_{s,t} p(s,t) \displaystyle\sum_a p(a,b=x_{s,a}^t|s,t)_{\sigma_{AB}}
\label{FUR2}
\end{eqnarray}
where the form of $p(a,b=x_{s,a}^t|s,t)_{\sigma_{AB}}$ in terms of the measurements on the bipartite state $\sigma_{AB}$ is given by
\begin{eqnarray}
p(a,b=x_{s,a}^t|s,t)_{\sigma_{AB}}= \displaystyle\sum_b V(a,b|s,t) \langle (A_s^a\otimes B_t^b)\rangle_{\sigma_{AB}}
\label{prob2}
\end{eqnarray}
where $A_s^a$ ($=\frac{(\mathcal{I}+(-1)^a A_s)}{2}$) is a measurement of the
observable $A_s$ corresponding to setting `s' giving outcome `a' at Alice's side; $B_t^b$ ($=\frac{(\mathcal{I}+(-1)^a B_s)}{2}$) is a measurement of the
observable $B_t$ corresponding to setting `t'  giving outcome `b' at Bob's 
side, and $V(a,b|s,t)$ is the winning condition given by
\begin{eqnarray}
V(a,b|s,t)&=& 1 \text{\phantom{xxxxx} iff $a\oplus b = s.t$} \nonumber \\
          &=& 0 \text{\phantom{xxxxx} otherwise}
\label{cond2}
\end{eqnarray}
Using  Eqs. (\ref{FUR2}), (\ref{prob2}), (\ref{cond2}) and taking $p(s,t)=p(s)p(t)=1/4$,   the expression of $P^{game}(\mathcal{S},\mathcal{T},\sigma_{AB})$ for the bipartite state $\sigma_{AB}$ is obtained to be  
\begin{eqnarray}
P^{game}(\mathcal{S},\mathcal{T},\sigma_{AB})=\frac{1}{2}(1+\frac{\langle\mathcal{B}_{CHSH}\rangle_{\sigma_{AB}}}{4})
\end{eqnarray}
where 
\begin{eqnarray}
\mathcal{B}_{CHSH}=A_0\otimes B_0+A_0\otimes B_1+A_1\otimes B_0-A_1\otimes B_1
\end{eqnarray}
 corresponds to the  Bell-CHSH operator$^{22,23}$. 
To characterize the allowed distribution under the theory, we need to know the maximum winning probability, maximized over all possible strategies for Alice and Bob. The maximum winning probability is given by
\begin{eqnarray}
P^{game}_{\max} = \max_{\mathcal{S},\mathcal{T},\sigma_{AB}} P^{game}(\mathcal{S},\mathcal{T},\sigma_{AB})
\label{maxgame}
\end{eqnarray}
The value of $P^{game}_{\max}(\mathcal{S},\mathcal{T},\sigma_{AB})$ allowed by classical physics is $\frac{3}{4}$ (as classically, the Bell-CHSH inequality  is bounded by $2$), by quantum mechanics is $(\frac{1}{2}+\frac{1}{2 \sqrt{2}})$ (due to the maximum violation of Bell inequality,  $\langle \mathcal{B}_{CHSH} \rangle =2\sqrt{2}$), and by no-signaling theories with maximum Bell violation ($\langle \mathcal{B}_{CHSH}\rangle=4$, that occurs for the PR-box$^{56}$ 
is $1$. The connection of
Eq.(\ref{cond2}) with the no-signalling constraint for the general case of a 
bipartite system was elaborated by Barrett et al.$^{57}$ 

The above description refers to the scenario when the two parties have no bias 
towards choosing a particular measurement. Nonlocality in the context of
biased games has been discussed in Ref.$^{30}$ using the fine-grained 
uncertainty relation.
In the particular game chosen$^{58}$ 
the biased game, the 
intention of Alice is to choose $A_{0}$ with probability 
$p$($0\leqslant p\leqslant 1$) and $A_{1}$ with probability $(1-p)$. Bob 
intends to choose $B_{0}$ and $B_{1}$ with probabilities 
$q$($0\leqslant q\leqslant 1$) and $(1-q)$,  respectively. The measurements 
and their outcomes are coded into binary variables pertaining to  an 
input-output process. Alice and Bob have binary input variables $s$ and $t$, 
respectively, and output variables $a$ and $b$, respectively. Input $s$ takes 
the values $0$ and $1$ when Alice measures $A_{0}$ and $A_{1}$, respectively. 
Output $a$ takes the values $0$ and $1$ when Alice gets the measurement 
outcomes $+1$ and $-1$, respectively. The identifications are similar for 
Bob's variables $t$ and $b$. Now, the rule of the game is that Alice and Bob's 
particles win (as a team) if their inputs and outputs satisfy 
\begin{equation}
a\oplus b = s.t
\end{equation}
where $\oplus$ denotes addition modulo $2$. Input questions $s$ and $t$ have 
the probability distribution $p(s,t)$ (for simplicity, $p(s,t)= p(s)p(t)$ where $p(s=0)=p$, $p(s=1)= (1-p)$, $p(t=0)= q$ and $p(t=1)= (1-q)$). 
The fine-grained uncertainty relation is now invoked.
The 
expression of $P^{game}$ is given by 
\begin{equation}
P^{game}(\mathcal{S},\mathcal{T},\rho_{AB})= \frac{1}{2}[1+ \langle CHSH(p,q)\rangle_{\rho_{AB}}]
\end{equation}
with $CHSH(p,q)= [pq A_{0}\otimes B_{0}+ p(1-q)A_{0}\otimes B_{1}+ (1-p)q A_{1}\otimes B_{0}-(1-p)(1-q)A_{1}\otimes B_{1}]$ being the form of CHSH-function after 
introducing bias. 

The maximum probability $P^{game}$ of winning the biased game was
obtained$^{30}$ by maximizing the function $\langle CHSH(p,q)\rangle$ for different 
theories. Such maximization was earlier performed in the literature for the
unbiased  scenario$^{59}$ 
and subsequently, 
for the biased case as well$^{58}$,  
in the latter case  by considering two halves of the ranges of
 the parameters $p$ and $q$. First, for the case of $p,q\geq 1/2$, 
the classical maximum is obtained using an extremal strategy where the values 
of all the observables are $+1$  giving the maximum value of the biased
CHSH-function to be 
$1-2(1-p)(1-q)$.
With this classical maximum, the winning probability is given by$^{30}$
\begin{equation}
P^{game}(\mathcal{S},\mathcal{T},\rho_{AB})|^{classical}_{maximum}= 1-(1-p)(1-q)
\end{equation}
This reduces to the value $\frac{3}{4}$ for the unbiased case when 
$p=q=\frac{1}{2}$. 
For the quantum strategy,
the parameter space is divided in two regions of [$p,q$] with the first region 
corresponding to $1\geq p\geq (2q)^{-1}\geq \frac{1}{2}$. Here 
$\langle CHSH(p,q)\rangle \leq 1-2(1-p)(1-q)$
 leads to
\begin{equation}
 P^{game}(\mathcal{S},\mathcal{T},\rho_{AB})|^{region}_1= 1-(1-p)(1-q)~~
\end{equation}
showing that the
upper bound is the same as achieved by classical theory. Thus,
quantum correlations (entanglement) offers no advantage  over  classical 
correlations  in performing the specified task in this region. 
However, in the  other region $1\geq (2q)^{-1}> p\geq \frac{1}{2}$
one gets the value
$\langle CHSH(p,q)\rangle \leq \sqrt{2}\sqrt{q^{2}+(1-q)^{2}}\sqrt{p^{2}+(1-p)^{2}}$
that is greater than the classical bound. So, the 
biasing parameters in this region enable discrimination among classical and 
quantum 
correlations. The upper bound of the fine-grained uncertainty relation  is in this case given by,
\begin{eqnarray}
P^{game}(\mathcal{S},\mathcal{T},\rho_{AB})&|&^{quantum}_{maximum}\nonumber\\
=\frac{1}{2}[1+\sqrt{2}\sqrt{q^{2}+(1-q)^{2}}&&\sqrt{p^{2}+(1-p)^{2}}]
\end{eqnarray}
 The extent
of non-locality that can be captured by the fine-grained
uncertainty relation is regulated by
the bias parameters.

The fine-grained uncertainty relation has been applied to study the nonlocality
 of tripartite systems, as well$^{29}$.
In this case a nonlocal retrieval game 
similar to CHSH-game for bipartite systems is considered, as follows.  Three
parties, Alice, Bob and Charlie 
receive respective binary questions `s', `t', and  `u' $\in\{0,1\}$ (corresponding to their two different measurement settings at each side), and they win the 
game if their respective outcomes (binary) `a', `b', and `c' $\in\{0,1\}$ satisfy certain rules. Three kinds of no-signaling 
boxes, known as full-correlation boxes have been considered, for which all one-party and two-party  
correlation in the system vanish$^{60}$. 
The
 game is won if their answers satisfy the set of rules,
either 
\begin{eqnarray}
a\oplus b \oplus c=s.t \oplus t.u \oplus u.s
\label{box1}
\end{eqnarray}
or 
\begin{eqnarray}
a\oplus b \oplus c=s.t \oplus s.u
\label{box2}
\end{eqnarray}
or else
\begin{eqnarray}
a\oplus b \oplus c=s.t.u
\label{box3}
\end{eqnarray}
All the above boxes violate the Mermin inequality$^{61}$, 
whereas the 
Svetlichny inequality$^{62}$ 
is violated only by the box given by Eq. (\ref{box1}) (known as the Svetlichny box). The winning probability of the game under a physical theory described by a shared tripartite state $\sigma_{ABC}$ (among Alice, Bob and Charlie) is given by
\begin{eqnarray}
&& P^{game}(\mathcal{S},\mathcal{T},\mathcal{U},\sigma_{ABC})\nonumber \\
&&=\displaystyle\sum_{s,t,u} p(s,t,u) \displaystyle\sum_{a,b} p(a,b,c=x_{s,t,a,b}^{(u)}|s,t,u)_{\sigma_{ABC}}
\label{FUR3}
\end{eqnarray}
where $p(s,t,u)$ is the probability of choosing the measurement settings `s' by Alice, `t' by Bob and `u' by Charlie, and  $p(a,b,c|s,t,u)_{\sigma_{ABC}}$ the joint probability of getting outcomes `a', `b' and `c' for corresponding settings `s', `t' and `u' given by
\begin{eqnarray}
&& p(a,b,c=x_{s,t,a,b}^{(u)}|s,t,u)_{\sigma_{ABC}}  \nonumber \\
&&  =\displaystyle\sum_{c}  V(a,b,c|s,t,u) \langle A_s^a\otimes B_t^b\otimes C_u^c\rangle_{\sigma_{ABC}}
\label{prob3}
\end{eqnarray}
where $A_s^a$, $B_t^b$ and $C_u^c$ are the measurements corresponding to setting `s' and outcome `a' at Alice's side, setting `t' and outcome `b' at Bob's side, and setting `u' and outcome `c' at Charlie's side, respectively; and $V(a,b,c|s,t,u)$ (the winning condition) is non zero ($=1$) only when the outcomes of Alice, Bob and Charlie are correlated by either of Eqs. (\ref{box1}), (\ref{box1}) or (\ref{box3}), and is zero otherwise. The maximum winning probability over all 
possible strategies (i.e., the choice of the shared tripartite state and 
measurement settings by the three parties) for any theory is given by
\begin{eqnarray}
P^{game}_{\max} = \max_{\mathcal{S},\mathcal{T},\mathcal{U},\sigma_{ABC}} P^{game}(\mathcal{S},\mathcal{T},\mathcal{U},\sigma_{ABC})
\end{eqnarray}
which is a signature of the allowed probability distribution under that theory.

The cases corresponding to classical, qauntum and no-signalling
theories with super-quantum correlations for the above different 
full-correlation boxes (rules of the nonlocal 
game) have been studied in Ref.$^{29}$.  
For the case of the winning condition given by Eq. (\ref{box1}), the expression of $P^{game}(\mathcal{S},\mathcal{T},\mathcal{U},\sigma_{ABC})$ for the shared tripartite state $\sigma_{ABC}$ is given by
\begin{eqnarray}
P^{game}(\mathcal{S},\mathcal{T},\mathcal{U},\sigma_{ABC})=\frac{1}{2} [1+\frac{\langle \textbf{S}_1\rangle_{\sigma_{ABC}}}{8}]
\end{eqnarray}
where
\begin{eqnarray}
\textbf{S}_1=&&A_0\otimes B_0\otimes C_0+A_0\otimes B_0\otimes C_1+A_0\otimes B_1\otimes C_0 \nonumber \\
&& +A_1\otimes B_0\otimes C_0-A_0\otimes B_1\otimes C_1-A_1\otimes B_0\otimes C_1\nonumber \\
&&-A_1\otimes B_1\otimes C_0-A_1\otimes B_1\otimes C_1
\end{eqnarray}
The value of $P^{game}_{\max}$ allowed in classical physics is $3/4$ which 
follows from the Svetlichny inequality$^{62}$ 
\begin{eqnarray}
\langle \textbf{S}_1\rangle_{\sigma_{ABC}} \leq 4
\end{eqnarray}
For the case of quantum physics, the 
maximum violation of the Svetlichny inequality is $4\sqrt{2}$ which occurs for
the GHZ-state$^{63}$.  
 The value of $P^{game}_{\max}$ allowed in quantum physics is $(\frac{1}{2}+\frac{1}{2 \sqrt{2}})$. For the 
case of the no-signalling theory, the algebraic maximum of the  Svetlichny inequality  is $8$, and the value of $P^{game}_{\max}$ in
this case
is $1$, corresponding to a correlation with maximum nonlocality.

It was found in Ref.$^{29}$ that none of the other two
 full corelation Mermin 
boxes (\ref{box2}) and (\ref{box3}) are
able to distinguish classical theory from quantum theory in terms of their
degree of nonlocality. The fine-grained uncertainty relation determines the
degree of nonlocality as manifested by the Svetlichny inequality
for tripartite systems corresponding to the wining
condition given by (\ref{box1}), in the same way as
it determines the nonlocality of bipartite systems manifested
by Bell-CHSH inequality. One is able to
differentiate the various classes of theories (i.e., classical
physics, quantum physics and no-signaling theories with
maximum nonlocality or superquantum correlations) by
the value of $P^{game}_{\max}$
for tripartite systems. A biased tripartite system had also been 
explored$^{30}$. However, it was observed using a bipartition model$^{64}$  
that there is a zone specified
by the biasing parameters where even the Svetlichny inequality
cannot perform the discrimination between various physical systems
based on their degree of nonlocality.

\section{Quantum memory}

In quantum information theory, an uncertainty relation in terms of entropy is 
regarded to be more useful than that in terms of standard deviation. 
 The uncertainty relating to the outcomes of observables 
is reformulated in terms of Shannon entropy instead of
standard deviation.
Entropic uncertainty relations for two observables in the context of discrete
variables was introduced by Deutsch$^{7}$.
An improved version was conjectured by Kraus$^{8}$, 
given by 
\begin{eqnarray}
\mathcal{H}(R)+\mathcal{H}(S) \geq \log_2 \frac{1}{c}
\label{EUR1}
\end{eqnarray}
and later proved by Maassen and Uffink$^{9}$. 
Here $\mathcal{H}(i)$ denotes the Shannon entropy of the probability 
distribution of the measurement outcomes of observable $i$ ($i\in\{R,S\}$) and 
$\frac{1}{c}$ quantifies the complementarity of the observable. For 
non-degenerate observables, $c= \max_{i,j}c_{i,j} = \max_{i,j} |\langle a_i|b_j\rangle|^2$, 
where $|a_i\rangle$ and $|b_j\rangle$ are eigenvectors of $R$ and $S$, 
respectively.

Using entanglement between the state of the observed system and another quantum 
system (memory), Berta et al.$^{10}$  have shown that the lower bound 
of entropic uncertainty  (given by Eq.(\ref{EUR1})) can be improved 
in the presence of quantum correlations.  The entropic uncertainty relation 
in the presence of quantum memory is given by$^{10}$
\begin{eqnarray}
\mathcal{S}(R|B)+\mathcal{S}(S|B) \geq \log_2 \frac{1}{c} + \mathcal{S}(A|B)
\label{EUR-QM}
\end{eqnarray}
where $\mathcal{S}(R|B)$ ($\mathcal{S}(S|B)$) is the conditional von Neumann 
entropy of the state given by
$\sum_{j} (|\psi_j\rangle\langle\psi_j|\otimes I)\rho_{AB}(|\psi_j\rangle\langle\psi_j|\otimes I)$,
with $|\psi_j\rangle$ being the eigenstate of observable $R  (S)$, and 
$\mathcal{S}(R|B)$ ($\mathcal{S}(S|B)$) quantifies the uncertainty corresponding to the measurement $R (S)$ on the system ``A" given information stored in the 
system ``B" (i.e., quantum memory).  $\mathcal{S}(A|B)$ quantifies the amount 
of entanglement between the quantum system possessed by Alice and the quantum 
memory possessed by Bob. 
For example, the sum of uncertainties of two measurement outcomes ($\mathcal{H}(R)+\mathcal{H}(S)$) for measurement of two observables $(R,S)$ on the quantum system (``A", possessed by Alice) can be reduced to $0$ (i.e., there is no uncertainty) if 
that system is maximally entangled with an another system, called quantum 
memory (``B", possessed by Bob). Here, Bob is able to reduce his uncertainty 
about Alice's measurement outcome with the help of communication from Alice 
regarding the choice of her 
measurement performed, but not its outcome.

Recently,
Coles and Piani$^{14}$  have  made the lower bound of entropic uncertainty in
the presence of quantum memory tighter.  Their modified form of the
entropic uncertainty relation is given by 
\begin{eqnarray}
\mathcal{S}(R_A|B)+\mathcal{S}(S_A|B) \geq c^{\prime}(\rho_{A}) + \mathcal{S}(A|B)
\label{EUR-QM_CP}
\end{eqnarray}
where $c^{\prime}(\rho_{A}) = \max\{c^{\prime}(\rho_A,R_A,S_A), c^{\prime}(\rho_A,S_A,R_A) \}$. $c^{\prime}(\rho_A,R_A,S_A)$ and $c^{\prime}(\rho_A,S_A,R_A)$ are defined by
\begin{eqnarray}
c^{\prime}(\rho_A,R_A,S_A) = \displaystyle\sum_i p^r_i \log_2 \frac{1}{\max_j c_{ij}} \nonumber \\
c^{\prime}(\rho_A,S_A,R_A) = \displaystyle\sum_j p^s_j \log_2 \frac{1}{\max_i c_{ij}}, 
\label{Complemetary_RS}
\end{eqnarray}
where $p^r_i = \langle r |\rho_{A}|r \rangle$ with $\sum_i p^r_i = 1$ and
$p^s_j =\langle s |\rho_{A}|s \rangle $ with $\sum_j p^s_j=1$.
Here, the uncertainty for the measurement of the observable $R_A$ ($S_A$) on
Alice's system by accessing the information stored in the
quantum memory with Bob  is measured by $\mathcal{S}(R_A|B)$ ($\mathcal{S}(S_A|B)$)
which is the conditional von Neumann entropy of the  state given by
\begin{eqnarray}
\rho_{R_A(S_A)B}&=&\sum_{j} (|\psi_j\rangle_{R_A(S_A)}\langle\psi_j|\otimes I)\rho_{AB}(|\psi_j\rangle_{R_A(S_A)}\langle\psi_j|\otimes I)\nonumber \\
&=&\sum_j p_j^{R_A(S_A)} \Pi_j^{R_A(S_A)}\otimes \rho_{B|j}^{R_A(S_A)},
\label{QState}
\end{eqnarray}
where $\Pi_j^{R_A(S_A)}$'s are the orthogonal projectors on the eigenstate $|\psi_j\rangle_{R_A(S_A)}$ of observable $R_A  (S_A)$, $p_j^{R_A(S_A)}=Tr[(|\psi_j\rangle_{R_A(S_A)}\langle\psi_j|\otimes I)\rho_{AB}(|\psi_j\rangle_{R_A(S_A)}\langle\psi_j|\otimes I)]$, $\rho_{B|j}^{R_A(S_A)}=Tr_A[(|\psi_j\rangle_{R_A(S_A)}\langle\psi_j|\otimes I)\rho_{AB}(|\psi_j\rangle_{R(S)}\langle\psi_j|\otimes I)]/p_j^{R_A(S_A)}$ and $\rho_{AB}$ is the state of joint system `$A$' and `$B$'.
In another work, Pati et al.$^{11}$  have extended the concept
of memory to include more general quantum correlations beyond entanglement. 
This leads to the improvement of the lower bound  given by
\begin{eqnarray}
\mathcal{S}(R_A|B)+\mathcal{S}(S_A|B) \geq && c^{\prime}(\rho_{A})  + \mathcal{S}(A|B) 
\label{EUR-QM_P} \\
&& + \max\{0,D_A(\rho_{AB})-C_A^M(\rho_{AB})\}, \nonumber
\end{eqnarray}
where the quantum discord $D_A(\rho_{AB})$ is given by$^{65}$ 
\begin{eqnarray}
D_A(\rho_{AB}) = \mathcal{I}(\rho_{AB})-  C_A^M(\rho_{AB}),
\label{QDis}
\end{eqnarray}
with $\mathcal{I}(\rho_{AB})$ ($=\mathcal{S}(\rho_A)+\mathcal{S}(\rho_B)-\mathcal{S(\rho_{AB})}$) being the mutual information of the state $\rho_{AB}$ which
contains the total correlation present in the state $\rho_{AB}$ shared between
the system $A$ and the system $B$, and the
classical information $C_A^M(\rho_{AB})$ for the shared state $\rho_{AB}$ (when Alice measures on her system) is given by
\begin{eqnarray}
C_A^M(\rho_{AB}) = \max_{\Pi^{R_A}}[\mathcal{S}(\rho_B) - \displaystyle\sum_{j=0}^1 p_j^{R_A} \mathcal{S}(\rho_{B|j}^{R_A}) ] 
\label{Cinf_M}
\end{eqnarray}

Experiments have demonstrated the
 effectiveness of reducing quantum uncertainty using
quantum memory, for the case of pure$^{66}$ 
as well as 
mixed$^{67}$ 
entangled states. 
For the purpose of experimental verification of inequality (\ref{EUR-QM}),
the entropic uncertainty is recast in the form of the sum of the Shannon
entropies $\mathcal{H}(p^R_d) + \mathcal{H}(p^S_d)$ when Alice and Bob measure the same observables
$R(S)$ on their respective systems and get different outcomes whose 
probabilities are denoted by $p^R_d(p^S_d)$, 
and $\mathcal{H}(p^{R(S)}_d) = -p^{R(S)}log_2p^{R(S)} -(1-p^{R(S)})log_2(1-p^{R(S)})$.  
Making use of Fano's 
inequality$^{68}$, 
it follows that $\mathcal{H}(p^R_d) + \mathcal{H}(p^S_d) \ge \mathcal{S}(R|B)+\mathcal{S}(S|B)$
which using  Eq.(\ref{EUR-QM}) gives \cite{Expt.1}
\begin{eqnarray}
\mathcal{H}(p^R_d) + \mathcal{H}(p^S_d) \ge  \log_2 \frac{1}{c} + \mathcal{S}(A|B)
\label{EUR-QM2}
\end{eqnarray}
The right hand side of the inequality (\ref{EUR-QM2}) can be
determined from the knowledge of the state and the measurement settings.
The entropic uncertainty relation has been used
for verifying the security of key distribution protocols$^{69}$.  
Devetak and Winter$^{70}$ 
derived that the amount of key $K$ that 
Alice and Bob are able to
extract per state  should always exceed the quantity $\mathcal{S}(R|E)
- \mathcal{S}(R|B)$, where the quantum state  $\rho_{ABE}$ is shared between 
Alice, Bob and the evesdropper Eve ($E$). Extending this idea
by incorporating the effect of shared quantum correlation between Alice and
Bob, Berta et al.$^{11}$ reformulated their relation (\ref{EUR-QM}),
in the form of $\mathcal{S}(R|E) + \mathcal{S}(R|B) \ge \log_2 \frac{1}{c}$
 enabling them to derive a new lower bound
on the key extraction rate, given by $K \ge \log_2 \frac{1}{c} - \mathcal{S}(R|B)- \mathcal{S}(S|B)$.

It has been recently realized that a
further improvement in the lower bound of entropic uncertainty is possible
using fine graining. 
A new form of the uncertainty relation 
in the presence of quantum memory was derived$^{16}$, in which the lower bound 
of entropic 
uncertainty corresponding to the
measurement of two observables is determined by fine-graining of the
possible measurement outcomes.  The fine-grained uncertainty 
relation$^{15}$, as discussed in the previous section, is here considered
in the context of
a quantum game played by Alice and Bob who share a two-qubit state $\rho_{AB}$ which is prepared by Alice.  Bob's qubit which he receives from Alice, 
represents the 
quantum memory.  Bob's uncertainty of 
the outcome of Alice's measurement of one of two incompatible observables (say, $R$ and $S$), is reduced with the help of fine-graining, when Alice helps 
Bob  by communicating her measurement choice of 
a suitable spin observable on her system. In this game Alice and Bob are 
driven by the requirement of minimizing the value of the quantity 
$\mathcal{H}(p^R_d) + \mathcal{H}(p^S_d)$ which forms the left hand side of the entropic 
uncertainty relation (\ref{EUR-QM2}). The minimization is over all 
incompatible measurement settings
such that $R \neq S$,
i.e., 
\begin{eqnarray}
\mathcal{H}(p^R_d) + \mathcal{H}(p^S_d) \ge \min_{R, S\neq R}[\mathcal{H}(p^R_d) + \mathcal{H}(p^S_d)]
\label{step1}
\end{eqnarray} 
To find
the minimum value, the choice of the variable $R$ was fixed
without the loss of generality to be $\sigma_z$ (spin measurement 
along the $z$-direction), and then the minimization was performed over the
other variable $S$. 
The uncertainty defined by the entropy $ \mathcal{H}(p^S_d)$ is minimum when
the certainty of the required outcome is maximum, corresponding to an
infimum value for the probability $p^S_d$.
In order to obtain the infimum value of $p^S_d$, the fine-grained
uncertainty relation was used in a form relevant to the 
game considered  where the
infimum value of the winning probability (corresponding to minimum uncertainty)
is given by 
\begin{eqnarray}
p_{\inf}^{S}= \inf_{S(\ne \sigma_z)}\displaystyle\sum_{a,b} V(a,b) Tr[(A_S^a\otimes B_S^b).\rho_{AB}], 
\label{inf-S}
\end{eqnarray}
with the winning condition $V(a,b)$ given by
\begin{eqnarray}
V(a,b) &=& 1 \textit{~~~~~ iff $a\oplus b=1$} \nonumber \\
       &=& 0 \textit{~~~~~ otherwise}.
\end{eqnarray}
with $A_S^a$ being a projector for observable $S$ with outcome `$a$', given by
$S^{\alpha}=\frac{I+(-1)^{\alpha} \vec{n}_{S}.\vec{\sigma}}{2}$ (and similarly for
$B_S^b$),
where $\vec{n}_{S}(\equiv \{\sin(\theta_{S}) \cos(\phi_{S}), \sin(\theta_{S}) \sin(\phi_{S}),$ $\cos(\theta_{S}) \} )$; $\vec{\sigma}\equiv \{\sigma_x,\sigma_y,\sigma_z\}$ are the Pauli matrices; $\alpha$ takes the value either $0$ (for spin 
up projector) or $1$ (for spin down projector). The above
winning condition proposed in Ref.$^{16}$ is different from the winning 
conditions
used in Refs.$^{15,29,30}$  for the purpose of capturing the nonlocality of
quantum systems. Here the fine-grained uncertainty
relation is to make
it directly applicable to the experimental situation of quantum
memory$^{66,67}$. 

The form of the entropic uncertainty relation obtained 
by fine-graining is given by$^{16}$ to be 
\begin{eqnarray}
\mathcal{H}(p^R_d)+\mathcal{H}(p^S_d) \geq  \mathcal{H}(p^{\sigma_z}_d) +
\mathcal{H}(p^S_{inf})
\label{FURQM1}
\end{eqnarray}
 The
value of $p^S_{inf}$ has been calculated for various quantum states
such a the Werner state, Bell-diagonal state and a state with maximally 
mixed marginals$^{16}$.
The above uncertainty relation (\ref{FURQM1})  is able to account for
the experimental results obtained for the case of maximally entangled
states$^{66}$ and mixed Bell-diagonal states$^{67}$. Moreover,
the limit set by (\ref{FURQM1}) prohibits the attainment
of the lower bound of entropic uncertainty$^{10}$ as defined by
the right hand side of  equation (\ref{EUR-QM})
 for the class of two-qubit states with maximally mixed marginals.

The uncertainty relation (\ref{FURQM1}) is independent of the choice of 
measurement settings as it 
optimizes the reduction of uncertainty
quantified by the conditional Shannon entropy over all possible observables.
Given a bipartite state possessing quantum correlations, inequality 
(\ref{FURQM1}) provides the fundamental limit to which uncertainty in the
measurement outcomes of any two incompatible variables can be reduced.
Since the uncertainty principle in its entropic form could be used
for verifying the security of key distribution protocols,
there exist ramifications of Eq.(\ref{FURQM1}) on the key extraction rate
in quantum key generation. It
is possible to obtain a tighter lower bound on the key rate$^{16}$ given by  $K \ge \log_2 \frac{1}{c} - \mathcal{H}(p^{\sigma_z}_d) + \mathcal{H}(p^S_{inf})$ when the
two parties involved in the protocol retain  data whenever they make the same choice
of measurement on their respective sides. The relation (\ref{FURQM1}) is the
 optimized lower 
bound of entropic uncertainty, which represents the ultimate limit to
which uncertainty of outcomes of two non-commuting observables can be
reduced by performing any set of measurements  in the presence of quantum 
memory.

\section{Conclusions}

In this article we have discussed various applications of different
versions of uncertainty relations. Much of the review presented
here deals with various formulations of entropic uncertainty 
relations$^{6,8,9,10,11,14,16}$ 
in different situations for the case of both discrete and continuous
variables. However, we have also briefly discussed the Heisenberg uncertainty
relation$^{1}$ and its Robertson-Schrodinger variant$^{3,4}$ in the 
context of two
specific applications, namely, demonstration of EPR-steering$^{18,19}$, 
and determination of
the purity of states$^{28}$, respectively. We conclude with a section-wise
summary of the main results discussed in this article, and a few possible future
directions of study.

We have discussed in Section II how the
Robertson-Schrodinger uncertainty relation
may be connected to the property of purity and mixedness
of  single and bipartite  qubit systems$^{28}$. The
uncertainty corresponding to the measurement of suitable observables
vanishes for pure states,  and is positive definite for mixed states. Using this
feature  a scheme was proposed to distinguish pure and mixed states 
belonging to the classes of  all single-qutrit states up
to three parameters, as well as several classes of
two-qutrit states, when prior knowledge of the basis is available$^{28}$. 
A possible implementation of the  proposed witnesses for detecting mixedness
here could be through techniques involving measurement of two-photon 
polarization-entangled modes for qutrits$^{71}$. 
Since the 
class of all pure states is not convex, 
the witnesses proposed for detecting mixedness do not arise from the 
separability criterion that holds for the widely studied entanglement 
witnesses$^{37}$,
 as well as the recently proposed teleportation witnesses$^{38}$,
and witnesses for absolutely separable states$^{39}$.
   However, a similar prescription of distinction of categories
of quantum states based on the measurement of expectation values of Hermitian
operators is followed. 

 In Section III a discussion of EPR steering$^{18,19,41}$ is presented
in the context of
continuous variable entangled states. 
Though entangled states form a strict subset of steerable states$^{21,45}$, 
several 
entangled
pure states fail to reveal steering through
the Reid criterion$^{19}$ for wide ranges of parameters.
Using the entropic uncertainty relation for continuous variables$^{6}$,
an entropic steering condition can be derived$^{43}$. Examples of various
non-Gaussian states for which entropic steering can be demonstrated, such as, 
the two-dimensional harmonic oscillator states,
the photon subtracted squeezed vacuum state, and the N00N state
have been studied$^{24}$.
 Steering with
such states may be demonstrated by computing the relevant conditional 
entropies using the
Wigner function whose non-Gaussian nature plays an inportant role.
These examples reiterate the fact that though Bell violation
guarantees steerability, the two types of quantum correlations are distinct
from each other. Moreover, the presence of quantum correlations in certain
class of states may be more easily detected through the violation of the 
entropic steering inequality compared to the violation of the Bell 
inequality$^{24}$. This could be useful for detecting and 
manipulating correlations in non-Gaussian states for practical purposes
in information processing and quantum metrology.

The relation between uncertainty and nonlocality is discussed in Section IV.
The connection between the  degree of nonlocality of the underlying physical 
theory and the fine-grained uncertainty relation has been proposed$^{15}$, as 
expressed in 
terms of the maximum 
winning probability of certain 
nonlocal games. A generalization of this connection  to the case of 
tripartite systems has been formulated$^{29}$. The 
fine-grained uncertainty relation determines the degree of nonlocality as 
manifested by the Svetlichny inequality$^{62}$ for tripartite systems in the same way 
as it determines the nonlocality of bipartite systems manifested by 
Bell-CHSH inequality$^{22,23}$. With the 
help of the fine-grained uncertainty relation, one is able to differentiate the 
various classes of theories (i.e., classical physics, quantum physics and no-signaling theories with maximum nonlocality or superquantum correlations) by the value of the maximum winning probability of the relevant retrieval game.  The  
fine-grained uncertainty relation$^{15}$ has been further employed$^{30}$  
to distinguish between 
classical, quantum and super-quantum correlations based on their strength of
nonlocality, in the context of biased games$^{58}$
involving two or three parties. Discrimination among the underlying theories 
with different degrees of nonlocality is in this case possible for a
specific range of the biasing parameters where quantum correlations offer the
advantage of winning the particular nonlocal game over classical correlations.
 Analytical generalizations to multiparty nonlocal 
games may further be feasible using such an approach$^{30}$.

Section V deals with the issue of entropic uncertainty relations for
discrete variables in the presence of quantum memory$^{10}$.
The optimized lower bound of 
entropic uncertainty  in the presence of quantum memory has been
derived$^{16}$
with the help of the
fine-grained uncertainty principle$^{15}$.  
 Since entropy (or uncertainty) is directly related to
probability, the analysis of fine-graining involves the minimization (or
maximization) of probability in order to minimize
uncertainty.  
In  measurements and communication involving
two parties, the lower bound of entropic uncertainty cannot fall below
the bound derived using fine-graining,  as is illustrated with several 
examples of pure and mixed states of discrete variables$^{16}$.  After
fine-graining the entropic uncertainty relation furnishes a fundamental 
limitation on the precision
of outcomes for measurement of two incompatible observables in the presence
of quantum memory. Implications on  the key rate for secure  key generation
is also discussed. Further work along this direction may be able to shed
light on the information theoretic resources entailed in the process of
fine-graining.

\section*{Acknowledgments}

ASM acknowledges support from the project SR/S2/LOP-08/2013
of DST, India.  TP acknowledges financial support from ANR 
retour des post-doctorants NLQCC (ANR-12-PDOC-0022- 01).

\end{document}